\documentclass[prb,twocolumn,showpacs,superscriptaddress,%
tightenlines]{revtex4}
\usepackage{amsmath,amsfonts,pifont,epsfig,color}
\usepackage{float}  

\newcommand{\bolS}{\mbox{\bf S}}
\newcommand{\bolT}{\mbox{\bf T}}
\begin{document}
\title{Low-lying excitations and magnetization process 
 of coupled tetrahedral systems}
\date{September 2001}
\author{K.~Totsuka}
\email[Present address: \\
Department of Physics, 
Aoyama Gakuin University ]{totsuka@phys.aoyama.ac.jp}
\author{H-J.~Mikeska}
\affiliation{Institut f\"{u}r Theoretische Physik, 
Universit\"{a}t Hannover, 
Appelstr. 2, 30167 Hannover, Germany}
\begin{abstract}
We investigate low-lying singlet and triplet excitations and the
magnetization process of quasi-1D spin systems composed of tetrahedral
spin clusters.  For a class of such models, we found various exact
low-lying excitations; some of them are responsible for the
first-order transition between two different ground states formed by
local singlets. Moreover, we find that there are two different kinds
of magnetization plateaus which are separated by a first-order
transition.
\end{abstract}
\pacs{75.10.Jm, 75.60.Ej, 75.40.-s}
\maketitle
\section{INTRODUCTION}
Recently, there has been substantial interest in the properties of 
strongly frustrated low-dimensional quantum spin systems such as 
$S=1/2$ Heisenberg models on the kagom\'{e} and pyrochlore 
lattices.  
In such systems, various ground-state phases, {\em e.g.} valence-bond 
crystals, RVB, long-range ordered phases etc, appear as we vary 
the control parameters.    
In particular, the possibility of a spin-liquid phase with 
unconventional singlet excitations would be 
interesting\cite{Lhuillier-S-F-00}.       

\begin{figure}[H]
\begin{center}
\epsfig{file=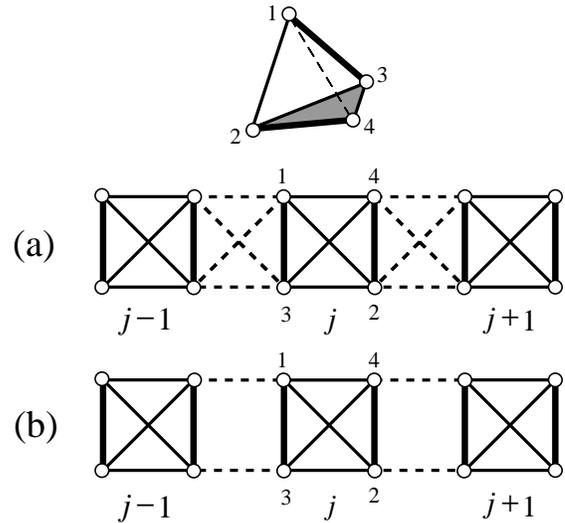,height=70mm}
\end{center}
\caption{Spin tetrahedron and two models considered 
in the text: symmetric model (a) and generalized model (b).}
\label{fig:tetra} 
\end{figure}

\begin{figure}[H]
\begin{center}
\epsfig{file=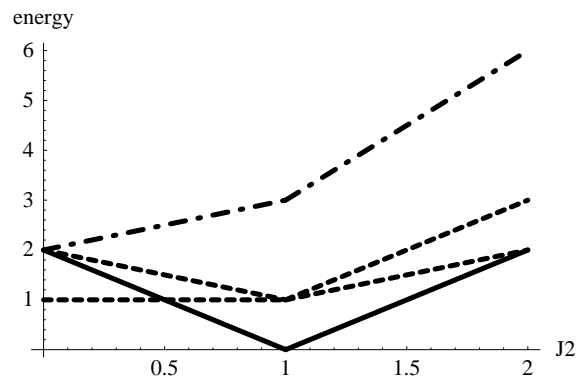,height=50mm}
\end{center}
\caption{Energy levels (in unit of $J_{1}$) 
of a single tetrahedron: 
singlet (solid), triplet (broken), and quintet (dot-dashed).  
Energy is measured from the ground state. Note that the ground state 
changes at $J_{1}=J_{2}$.}
\label{fig:single-levels}
\end{figure}

In the following, we consider 
a simple model system built up 
from tetrahedral clusters of $S=1/2$ 
(we label each tetrahedron by Latin indices $j$) with the Hamiltonian 
(see Fig.\ref{fig:tetra})  
\begin{align}
{\cal H}_{0} &= \sum_{j}J_{1}\left( 
\bolS_{j,1}{\cdot}\bolS_{j,3}+\bolS_{j,2}{\cdot}\bolS_{j,4} 
\right)    \notag \\
& + \sum_{j}J_{2}\left( 
\bolS_{j,1}{\cdot}\bolS_{j,2}+\bolS_{j,2}{\cdot}\bolS_{j,3}
+ \bolS_{j,3}{\cdot}\bolS_{j,4}+\bolS_{j,4}{\cdot}\bolS_{j,1} 
\right) \;  . 
\end{align}
These tetrahedra form a chain-like structure and 
interact with each other by the following coupling: 
\begin{align}
 {\cal H}_{1} &= \sum_{j} \bigl[ J_{3} \left( 
\bolS_{j,2}{\cdot}\bolS_{j+1,3}+\bolS_{j,4}{\cdot}\bolS_{j+1,1} 
\right)  \notag \\    
& + J_{4} \left( 
\bolS_{j,2}{\cdot}\bolS_{j+1,1}+\bolS_{j,4}{\cdot}\bolS_{j+1,3}
\right) \bigr] \; . 
\end{align}
For $J_{2}=J_{3}=J_{4}$, this model reduces to a known model of 
a frustrated spin ladder with diagonal couplings (`generalized Bose-Gayen 
model' \cite{Gelfand-91,Bose-G-93}).   
The choice $J_{3} = J_{4}\neq J_{2}$ ({\em symmetric model}, hereafter) 
introduces explicit dimerization in the leg direction.    
As can be seen in Fig.~\ref{fig:single-levels}, 
an assembly of decoupled tetrahedra described by ${\cal H}_{0}$ has 
singlet modes with energies much lower than the singlet-triplet gap.  
Therefore, the model with $J_{1},J_{2} \gg J_{3},J_{4}$ would 
provide a good starting point to study the unconventional  
properties mentioned above.   

The study of the model described by the Hamiltonian 
${\cal H}_{\text{tot}}\equiv {\cal H}_{0}+{\cal H}_{1}$ 
was inspired by the discovery of the tellurate  
materials \cite{Johnsson-T-M-M-00} 
$\text{Cu}_{2}\text{Te}_{2}\text{O}_{5}\text{X}_{5}$ (X= Cl or Br)   
and the phase diagram, triplet excitations, and the optical spectrum 
of this model have been investigated quite recently\cite{Brenig-B-01} 
using both numerical diagonalization and  
the bond-operator mean-field approximation.  
In these materials, $\mbox{Cu}^{2+}$ ions form $S=1/2$ tetrahedra, 
which are connected with each other by Te-O coordinations.  
Although the coupling between the tetrahedra is not so simple, 
the crystal structure\cite{Johnsson-T-M-M-00} ($\text{P}\bar{4}$) 
suggests that 
the model Hamiltonian ${\cal H}_{\text{tot}}$ with $J_{3}=J_{4}$  
is one of the simplest candidates to describe the tellurates 
(the chain axis is parallel to $c$-axis of the tellurates).  
On the basis of experimental results obtained 
for static magnetic properties \cite{Johnsson-T-M-M-00} 
and for Raman scattering \cite{Lemmens-01}, it was argued that 
the parameters $J_{1} \approx J_{2}$ and $J_{3} = J_{4} \ll J_{1}$ 
may be appropriate for the two compounds.   
We will also include some results  
for a more general case with $J_{3} \neq J_{4}$ 
(hence we call it {\em generalized model} in the following)  
which may clarify to what extent our results are general.   

\begin{figure}[H]
\begin{center}
\epsfig{file=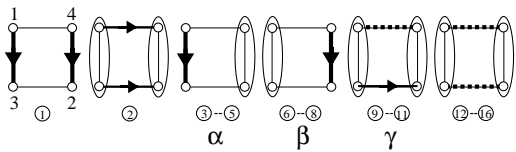,height=20mm}
\end{center}
\caption{Eigenstates of a single tetrahedron. Singlets: 1 (SD) and 
2 (S1D), 
triplets: $\alpha=(3,4,5)$, $\beta=(6,7,8)$, and 
$\gamma=(9,10,11)$, 
quintet: (12,13,14,15,16).  Arrows denote singlets and 
dashed lines and ovals triplets. 
In the dimer picture, `2', `9'-`11', and `12'-`16' can be 
viewed as 2-triplet bound states.}
\label{fig:single-states}
\end{figure}

Our aim here is to investigate (i) general excited states and their 
relation to the ground-state phase transitions and (ii) the magnetization 
plateaus.   In what follows, we will use eigenstates of the four $S=1/2$ 
spins on a single tetrahedron in the notation as shown in 
Fig.\ref{fig:single-states}.    
The ground state of the chain of non-interacting tetrahedra
$J_{3}=J_{4}=0$ is obtained as a sequence of tetrahedra in state `1'
for $J_{1}>J_{2}$ and as that of tetrahedra in state `2' for
$J_{1} < J_{2}$. 

We conclude this introduction with a few comments on the ground 
state phase diagram for the symmetric model (including the coupling 
$J_{3}=J_{4}$) which was investigated already 
in Ref.~\onlinecite{Brenig-B-01}.  
The main purpose of this part consists not in obtaining the 
phase diagram itself but in demonstrating how the low-lying singlets 
dictate the transition.   
To discuss the structure of the phase diagram from 
the viewpoint of the singlet spectra, 
we use an effective Hamiltonian derived by paying particular 
attention to low-energy singlets.   
This approach gives a simple and clear picture of the transition 
between singlet phases and will be useful also for a discussion 
of the low-energy singlet dynamics.     

The classical model ($S \nearrow \infty$) has two different kinds of  
antiferromagnetic phases separated by a line $J_{1}=J_{2}+J_{3}$.  
Along the transition line, the ground state exhibits a huge degeneracy, 
which is reminiscent of what occurs to the classical 
pyrochlore antiferromagnets\cite{Lhuillier-S-F-00}. 
As was already pointed out in Ref.~\onlinecite{Brenig-B-01}, 
the symmetric model ($S=1/2$) preserves  
the basic property\cite{Xian-95} of the generalized Bose-Gayen model: 
the total spins $\mbox{\bf S}_{j,1}+\mbox{\bf S}_{j,2}$   
and $\mbox{\bf S}_{j,3}+\mbox{\bf S}_{j,4}$ 
on individual dimer bonds ($J_{1}$) are 
well-defined quantum numbers.   
This allows a simple classification of the eigenstates of 
${\cal H}_{\text{tot}}$ 
by specifying these quantum numbers for all $2N$ ($N$: the number of 
tetrahedra) dimer bonds.   
When all dimers are occupied by triplets, the Hamiltonian 
${\cal H}_{\text{tot}}$ reduces to an effective spin-1 chain%
\cite{Gelfand-91}: 
\begin{equation}
{\cal H}^{S=1} = J_{2}\sum_{k\text{: odd}}\bolT_{k}{\cdot}\bolT_{k+1}
+ J_{3}\sum_{k\text{: even}}\bolT_{k}{\cdot}\bolT_{k+1}  \; ,
\label{eqn:spin-1chain}
\end{equation}
where $\bolT_{k}$ denotes an effective spin-1 operator 
$\bolS_{1,\frac{k+1}{2}}+\bolS_{3,\frac{k+1}{2}}$ (for $k$: odd) 
or $\bolS_{2,\frac{k}{2}}+\bolS_{4,\frac{k}{2}}$ (for $k$: even).    
On the other hand, interactions $J_{2}$ and $J_{3}$ 
effectively vanish when all dimer bonds are occupied by singlets.  

It was argued in Ref.~\onlinecite{Brenig-B-01} 
that the quantum phase diagram contains three different phases: 
rung-dimer (RD), Haldane\cite{Haldane-83} (H), 
and spin-1 dimerized (S1D, or plaquette singlet) 
phase\cite{Yamamoto-95,Totsuka-N-H-S-95};   
the first one is characterized by the formation of local singlets on 
dimer ($J_{1}$) bonds.  On the other hand, 
the latter two are phases of the model 
${\cal H}^{S=1}$ (eq.(\ref{eqn:spin-1chain})) 
and are distinguished\cite{Totsuka-N-H-S-95}  
according to whether the string-order parameter\cite{denNijs-R-89} 
is vanishing or not.  
Note that we can use a variational argument to show {\em rigorously} 
that RD phase actually 
realizes {\em at least} for $J_{1}>J_{2}+J_{3}$ ($J_{2},J_{3}>0$)
in the sense that no admixture of singlets `1' and `2' occurs.   

To investigate the effect of the coupling $J_{3}$ between 
tetrahedra on low-lying singlets analytically, 
we derive an effective Hamiltonian acting on the $2^{N}$-dimensional 
subspace spanned by two nearly degenerate states `1' and `2'.   
At `site'-$j$ (i.e. $j$-th tetrahedron), we define an Ising spin 
with a value +1 ($-1$) when the tetrahedron is in state 
`2' (`1').   
The coupling between tetrahedra is taken into account by 
degenerate perturbation theory and 
the resulting effective Hamiltonian is given by a ferromagnetic  
Ising chain in an external field: 
\begin{equation}
{\cal H}_{\text{Ising}} = J_{\text{IM}} \sum_{j} 
\sigma_{j}\sigma_{j+1} -h_{\text{IM}} \sum_{j} \sigma_{j} \; .
\label{eqn:effective1}
\end{equation}
with 
\begin{equation}
J_{\text{IM}} = -\frac{J_{3}^{2}}{6J_{2}} \qquad , \qquad 
h_{\text{IM}} = J_{2}-J_{1}+\frac{J_{3}^{2}}{3J_{2}}
\;  .
\end{equation}  
We have different ground states according to the sign of 
the effective magnetic field $h_{\text{IM}}$:  
For $h_{\text{IM}}<0$ the Ising spins align downward and 
the `rung dimer (RD)' phase realizes, while a positive value 
of $h_{\text{IM}}$ makes Ising spins point upward 
to form the `spin-1 dimer (S1D)' phase.  
The condition $h_{\text{IM}}=0$ 
determines the line of first-order transition between the RD- and the 
S1D phase.  
The mechanism of the transition will be discussed in the next section 
from the viewpoint of the excitation spectra.  
In Fig.\ref{fig:phase-diag1}, we show the transition line 
obtained above by a solid line.   
This result implies that at the symmetric point $J_{2}/J_{1}=1$ 
a small perturbation $J_{3}$ resolves the huge degeneracy and 
S1D phase is selected as a unique, spin-singlet ground state.   

The explicit form of the ground state is known analytically 
in the whole RD phase (all tetrahedra in state `1') and 
only on the line $J_3=0, J_2 > J_1$ in the S1D phase 
(all tetrahedra in state `2'). In the remaining 
part of the phase diagram, the ground state is known numerically 
from the study of $S=1$-chains with alternating exchange 
\cite{Yamamoto-95,Totsuka-N-H-S-95}.  
In particular the separation line between the Haldane  
and the S1D phase is given by 
$J_3/J_2 \approx 0.6$, where a second-order transition described by 
the $\theta=\pi$ O(3) non-linear sigma model\cite{Haldane-83} occurs. 
The quantum critical point, where 
all three phases meet and the gap of the first order transition
disappears, is of particular interest, but will have to be treated
beyond perturbation theory.
In section \ref{section:summary}, we briefly discuss the effect of 
{\em interchain} couplings in conjunction with three-dimensional 
ordering.   

As an independent check of our method, 
we also determined the RD-S1D boundary by adapting numerical data of 
Ref.~\onlinecite{Totsuka-N-H-S-95} 
(open squares in Fig.~\ref{fig:phase-diag1}. Note that this is
essentially the same as that given in Ref.~\onlinecite{Brenig-B-01}).  
The resulting ground-state phase diagram is shown in 
FIg.~\ref{fig:phase-diag1} as a function of  
$J_{2}/J_{1}$ and $J_{3}/J_{1}$ with $J_{1}$ as the energy unit. 
The merit of this representation is that the phase diagram clearly 
exhibits the symmetry under the exchange of $J_{2}$ and $J_{3}$.    
In the inset, we also show the same data in the 
$(J_{3}/J_{2},J_{1}/J_{2})$-plane, the parametrization used in 
a recent literature\cite{Brenig-B-01}, in order to facilitate 
a comparison.   Although we used a completely different method,   
the result is consistent with the one obtained by the bond-operator 
mean-field approach\cite{Brenig-B-01}.  
The inclusion of higher-order terms hardly changes the boundary 
and we may expect that the convergence of our calculation is good.    
We would like to stress here that our simple effective Hamiltonian 
(\ref{eqn:effective1}) not only yields a fairly good result\cite{Note-PD} 
but also gives us a clear picture of the transition 
as will be described in the next section.  

For the generalized model, the phase diagram is not  
symmetric and the relation to spin-1 chains is no longer useful.  
For small $J_{3}$ (or equivalently, $J_{4}$), however, the mapping 
to the Ising chain (\ref{eqn:effective1}) goes in a similar manner 
to give 
\begin{align}
J_{\text{IM}}&= -\frac{(J_{1}-3J_{2})^{2}J_{3}^{2}}{48J_{1}
(2J_{2}-J_{1})J_{2}} \quad , \notag  \\ 
h_{\text{IM}} &= J_{2}-J_{1} + 
\frac{(2J_{1}-3J_{2})(J_{1}-3J_{2})}{24J_{1}J_{2}(J_{1}-2J_{2})}J_{3}^{2}
\;  . 
\end{align}
Setting $J_{1}=J_{2}$, we obtain a negative value 
$h_{\text{IM}}=-J_{3}^{2}/(12J_{1})$.   
Contrary to the previous case (symmetric model), this implies that 
for perfect ($J_{1}=J_{2}$) tetrahedra the inter-tetrahedron 
coupling selects not the S1D ground state but the RD one.  
The results obtained in this section lead us to conclude that 
{\em if the value} $J_{1}=J_{2}$  
{\em is reliable for the real compounds} 
$\mbox{Cu}_{2}\mbox{Te}_{2}\mbox{O}_{5}\mbox{X}_{2}$ (X=Cl, Br), 
they should be treated in the S1D phase.    

\section{Excitations}
Because of the fully frustrated tetrahedral structure, there are
various types of singlet and triplet excitations.  Among them, 
singlet excitations within the triplet gap are  
of particular interest 
\cite{Mambrini-T-M-99,Waldtmann-98,Ramirez-H-W-00}.  
Main results of this section are summarized  
in Figs. \ref{fig:low-lying1}, \ref{fig:low-lying2} and Table I.    

\subsection{Excitations in the rung-dimer phase}
\subsubsection{Singlet excitations}
As can be easily seen from Fig.\ref{fig:single-levels}, 
low-lying singlet degrees of freedom do exist in the neighborhood of 
the point $J_{1}=J_{2}$ which corresponds to perfect tetrahedra.  
The lowest singlet in the rung-dimer (RD) phase is created 
by promoting one of 
the RD tetrahedra to an S1D singlet (state `2' 
in Fig.\ref{fig:single-states}).  
For the symmetric model, this is an exact eigenstate  
with energy $2(J_1-J_2)$ since the interaction Hamiltonian 
${\cal H}_1$ annihilates this state.    For an
estimate of the interaction effects ($\propto J_3$), the effective
Hamiltonian (\ref{eqn:effective1}) can be used and leads to the
results: 
\begin{align}
\Delta_{\text{RD}}^{\text{sing}} &= 2(J_{1}-J_{2}) \qquad  
\text{for the {\em symmetric model}} , \notag \\
\Delta_{\text{RD}}^{\text{sing}} 
&= 2(J_{1}-J_{2}) + \frac{J^{2}_{3}}{4}
\left( \frac{3}{4J_{1}}-\frac{1}{4J_{2}} \right) \notag \\
& \qquad \qquad \phantom{2(J_{1}-J_{2})} 
\text{for the {\em generalized model}} \; .   \label{eqn:RD-sing}
\end{align}
Alternatively this state can be viewed as a singlet bound state 
made up of two dimer triplets in the dimer region ($J_{1}\gg J_{2}$).   
The expression for the generalized model  
(second eq. of (\ref{eqn:RD-sing}))  
is, of course, not exact,
but we can show, at least in a perturbative sense, that this excitation 
is completely localized (i.e. dispersionless)  
also in the generalized model.   

What is more interesting is that there exist multiparticle 
bound states which is given by $n(\geq 2)$ successive S1D singlets; 
ferromagnetic interaction in the effective Hamiltonian provides 
the attraction between these particles (the binding energy is 
$-4(n-1)|J_{\text{IM}}|$).  
For both symmetric- and generalized models,  
the total energy of this kind of bound states is given by 
\begin{equation}
\Delta_{n\text{-bound}}^{\text{RD}} =  
\Delta_{\text{RD}}^{\text{sing}} +2(n-1) |h_{\text{IM}}|  \; .  
\end{equation}
Thus, when the phase boundary is approached from the RD side
($h_{\text{IM}} \rightarrow -0$), all these bound states collapse onto
the lowest singlet (while a small gap of order $J_{1}-J_{2}$ remains between
the singlet ground state and the lowest singlet excitation--%
see Fig.\ref{fig:low-lying1}).  
This collapse triggers phase separation and leads to the first-order
transition from singlet RD to singlet S1D.  
In particular, the largest one with the gap 
$\Delta^{N\text{-bound}}_{\text{RD}}=2N|h_{\text{IM}}|$ hits 
the ground state at the transition point and after the transition 
these huge bound states constitute low-lying excitations 
of the new phase.   Contribution of these low-lying singlets to 
such physical quantities as specific heat can be calculated by using 
the solvable Hamiltonian ${\cal H}_{\text{Ising}}$.    

On top of them, there are several singlet bound states composed of 
two {\em triplet} tetrahedra.  For example, a singlet combination of 
two triplet tetrahedra (state `$\alpha$' and `$\beta$') connected 
by a weak ($J_{3}$) link has an exact energy $2J_{1}-2J_{3}$, 
which lies between the two-triplet threshold $2J_{1}$ and the elementary 
singlet $\Delta_{\text{sing}}^{\text{RD}}$.    

\subsubsection{Triplet excitations}
In this subsection, we discuss several exact magnetic excitations 
for the symmetric model.  
Although most of the following results hold 
also for the generalized model, 
the excited states are no longer exact.  

The simplest such excitation is an immobile dimer triplet excitation
created by replacing one of the singlet (state `1') tetrahedra by a
tetrahedron in state `$\alpha$' or `$\beta$'.  
The exact excitation energy is
$J_{1}$, independent of $J_{2}$ and $J_{3}$. These are nothing but 
dimer-triplet excitations of the standard two-leg ladder with
Bose-Gayen type couplings.  

Another triplet with energy $2J_{1}-J_{2}$ can be created by promoting
one tetrahedron from state `1' to state `$\gamma$',  
composed of two dimer triplets.   
In the dimer limit ($J_{1} \gg J_{2},J_{3}$), this state
can be viewed as a triplet bound state made up of two dimer triplets
(with binding energy $-J_{2}$). If two dimer triplets are bound on a
weak ($J_{3}$) link, then they form another exact bound state with
energy $2J_{1} - J_{3}$, corresponding to adjacent tetrahedra in
states `$\alpha$' and `$\beta$'.   
By a logic similar to that used in the previous 
subsection (section II.A.1), we can show that these triplets are 
completely localized (i.e. dispersionless) even if we relax the condition 
$J_{3}=J_{4}$.      

\subsection{Excitations in the $S=1$ dimer phase}
\subsubsection{Singlet excitations}
Starting from the ground state of the S1D phase, elementary singlet
excitations for the symmetric model are obtained by changing some
tetrahedra to state `1'. Since the intertetrahedra coupling ${\cal H}_{1}$ 
annihilates links with at least one tetrahedron in state `1' on their 
edges, this change results in a sequence of finite $S=1$ chains. 
The excitation energy, however, cannot  
be calculated analytically because of quantum fluctuation coming 
from $S=1$ segments.   
For the lowest singlet
excitation, obtained for one tetrahedron in state `1', the energy is
obtained in second-order perturbation as (the second-order correction 
in the following expression is the contribution from the open ends 
of two $S=1$ chains)
\[ \Delta_{\text{S1D}}^{\text{sing}}=
2(J_{2}-J_{1})+4J^{2}_{3}/(3J_{2});  .  \]
We have $\Delta_{\text{S1D}}^{\text{sing}}>0$ in the S1D phase.   
It is easy to verify that $\Delta_{\text{S1D}}^{\text{sing}}$ 
coincides with the first equation of (\ref{eqn:RD-sing}) at the 
transition point $|h_{\text{IM}}|=0$.   

As in the case of the RD phase, we can consider 
several multiparticle states made up of state `1' tetrahedra  
(both scattering states and bound states).   
Among them, the most important is an $n$-particle 
bound state.   
The energy of immobile $n$-particle bound states can be 
calculated in perturbation theory as 
\[ \Delta_{\text{S1D}}^{\text{sing}} +
2(n-1)h_{\text{IM}} \; .  \]
Again, when $h_{\text{IM}}\approx 0$, the binding energy 
is so large that all multiparticle bound states have the
same energy $\Delta_{\text{S1D}}^{\text{sing}}$.  
This is the origin of instability towards the first-order transition 
from the S1D to the RD.  It is important to note that 
all the above excitations are {\em not} included in the effective 
spin-1 chain, but correspond to an internal degree of freedom 
on a pair of dimer spins.   

\subsubsection{Triplet excitations}

Two different types of triplet excitations exist.  
The first one is essentially a single tetrahedron  
in state-`$\alpha$'(`$\beta$') in the S1D background;  
it creates a single dimer singlet 
in the sea of dimer triplets and is {\em not} included in the usual 
`$S=1$ chain'.  This is highly localized because an `unpaired' spin-1 
object appearing at the edge of the effective $S=1$ chain (with an odd 
number of effective spins) can hardly move   
due to strong dimerization\cite{note-1}. 
The energy is given by 
\[ \Delta^{\text{triplet}}_{\text{type-1}}
=-J_{1}+2J_{2} +\mbox{O}(J_{3}^{3})  \; .  \]

Triplet excitations of the second type are contained in  
the excited states of 
`$S=1$ chain' and are obtained by promoting a single tetrahedron to
state-`$\gamma$'.   
The excitation discussed in Ref.~\onlinecite{Brenig-B-01}
by a mean-field approximation etc. is of this type.   
Contrary to the `$\gamma$' tetrahedron in the RD phase,     
this magnon excitation can propagate freely due to the `background' 
of dimer triplets and the dispersion is given by
\begin{align}
&\omega^{\text{triplet}}_{\text{type-2}}(q) \notag \\
& \qquad  = \left( J_{2}+\frac{8J^{2}_{3}}{27J_{2}} \right) 
-\left( \frac{4J_{3}}{3}+\frac{2J^{2}_{3}}{3J_{2}} \right) 
\cos q 
-\frac{4J^{2}_{3}}{9J_{2}} \cos 2q  \;  .
\end{align}
Note that this has a relatively large bandwidth 
of $0.92J_{1}$ (for $J_{1}=J_{2}$ and $J_{3}/J_{1}=0.3$).  
This fact shows that the effect of intertetrahedra couplings on 
low-lying excitations is drastically different according to 
the spin background.  This difference may be crucial also in 
considering possible scenario of three-dimensional ordering 
(see section\ref{section:summary}).  

Furthermore, by solving a two-magnon 
problem explicitly\cite{Mattis-book}, we found singlet- and triplet 
bound states around $q=\pi$, whose binding energies are a few percent 
of the triplet gap.  
The singlet one agrees qualitatively with that pointed out in 
Ref.\onlinecite{Brenig-B-01} in conjunction with the Raman spectrum.  
The detail will be reported elsewhere.   

Because of the large bandwidth, a crossing between the type-2 mobile 
triplet and the lowest gapped singlet (state-`1') occurs at relatively 
small value of $J_{2}/J_{1}$ (1.47 for $J_{3}/J_{1}=0.3$).  
For $0.5 < J_{2}/J_{1} < (J_{2}/J_{1})_{\text{c2}}$, 
the system is {\em unusual} in the sense that 
the ground state is dimer-like or 
$S=1$-like while the singlet-triplet gap is filled with  
many low-lying singlets.   
For larger values of $J_{2}$,  the tetrahedral chain is equivalent to a
standard $S=1$ chain not only for the ground state but also for low-lying 
excitations.   

\begin{figure}[b]
\begin{center}
\epsfig{file=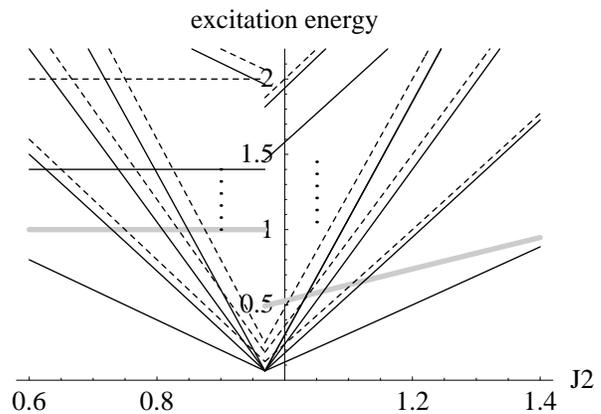,height=55mm}
\end{center}
\caption{Several low-lying singlets (in unit $J_{1}$) obtained by 
perturbation expansion in $J_{3}$ (some of them are exact) 
as a function of $J_{2}$ ($J_{3}$ is fixed).  
We chose $J_{3}=0.3 J_{1}$ (hence first-order transition occurs 
at $J_{2}/J_{1}=j_{\text{c}}=0.96904\ldots$).  
Elementary singlet and bound states are shown by solid line, while 
scattering states are shown by broken lines.   
Note that we show only a part of the entire singlet spectrum 
and actually a gap between the singlet ground state and the lowest 
triplet (shown by gray lines) is filled up with singlets composed 
of the elementary singlet discussed in section II.A.1.}
\label{fig:low-lying1}
\end{figure}

\begin{figure}[H]
\begin{center}
\epsfig{file=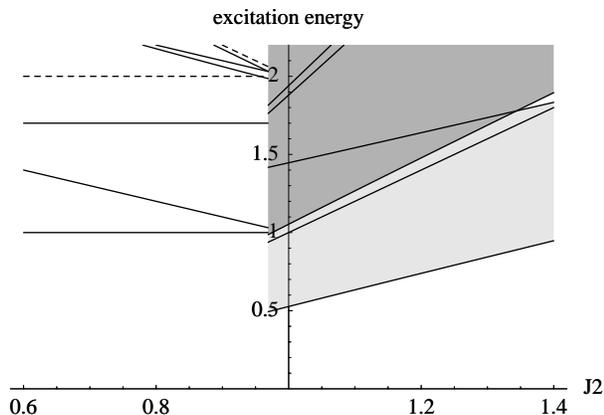,height=55mm}
\end{center}
\caption{Several low-lying triplets (in unit $J_{1}$) obtained by 
perturbation expansion in $J_{3}$ (some of them are exact) 
as a function of $J_{2}/J_{1}$.  
Parameters are the same as in Fig.\ref{fig:low-lying1}.  
1-triplet band and 2-triplet continuum are shown by light- and 
dark gray regions. Note that the dispersion of the lowest 
triplet suddenly changes at the transition since the ground 
states on both sides are completely different.}
\label{fig:low-lying2}
\end{figure}

We summarize the results obtained in this section  
in Fig.\ref{fig:low-lying1}, \ref{fig:low-lying2} and Table I.   
In Fig.\ref{fig:low-lying1}, we can see clearly the collapse of 
a huge number of singlet states mentioned above.   
Among them, the longest one (which is not shown here) comes down 
from infinitely high energies and hits the RD ground state at 
$J_{2}/J_{1}=(J_{2}/J_{1})_{\text{c1}}$.     
In Fig.\ref{fig:low-lying2}, some triplet branches appear 
discontinuous at the transition.  
This is because the corresponding state acquires 
an extensive dispersion when the critical  
value $J_{2}=J_{\text{2c}}$ is crossed from the RD side.   
The difference between a given type of excitation in a `background' 
of states `1' and `2' respectively , i.e. the discontinuity 
appearing in Fig.\ref{fig:low-lying2}, vanishes as the coupling 
$J_{3}$ decreases.   
For example, the level corresponding to the gap 
$\Delta_{\text{triplet}}^{\text{type-1}}$ has  
a discontinuity\cite{note-2} of the order $J_{3}^{2}$.      

\section{Magnetization Process}
In this section, we investigate the magnetization process of the
tetrahedral chain for fields up to the saturation field, paying
particular attention to the magnetization plateaus which appear at
$m^{z}/m_{\text{sat}}=1/2$ and are related to two different quantum
states. Since the real materials have $J_{1}\approx J_{2} \approx 40$
K, these plateaus may be detected in high-field magnetization
measurements.

The appearance of two different types of plateaus at
$m^{z}/m_{\text{sat}}=1/2$ is apparent already in the limit of
isolated tetrahedra (i.e. $J_3=0$); for $J_2 < J_1$ this plateau
occurs for magnetic fields $J_1 < H < J_1 + J_2$ and with the system
either in state `3' or in state `6' (type I plateau),  
whereas for $J_2 > J_1$ the plateau occurs for magnetic fields  
$J_2 < H < 2 J_2$ with the system in state `9' (type II plateau).

For a discussion of the magnetization process in the presence of 
interaction ($J_3 \ne 0$), we start by considering the dimer limit $J_{1}
\gg J_{2,3}$ where the plateau of type I is realized.  It is
well-known that the magnetization process in this limit can be reduced
to that of an effective $s=1/2$ model \cite{Totsuka-98}   
by regarding a triplet on a  
dimer bond as an upward spin ($s^z = +\frac{1}{2}$ ) and,
correspondingly, a singlet as downward spin ($s^z = -\frac{1}{2}$).  
The effective Hamiltonian obtained in this way is
\begin{widetext}
\begin{align}
{\cal H}_{\text{eff-1}} &= 
\sum_{(i,i+1)\in \text{between tetra.}} 
\left[ J_{xy}(s_{i}^{x}s_{i+1}^{x}+ s_{i}^{y}s_{i+1}^{y}) 
+ J_{zz}s_{i}^{z}s_{i+1}^{z} \right]  
 + J_{2} \sum_{(j,j+1) \in \text{tetra}} s^{z}_{j}s^{z}_{j+1} 
\notag \\ 
&  \qquad \qquad 
- (H-J_{1}-\frac{1}{2}J_{zz}-\frac{1}{2}J_{2})\sum_{k}s^{z}_{k} \; ,
\end{align}
where $J_{xy}$ and $J_{zz}$ are 
\begin{equation}
J_{xy}= \begin{cases} 0 & \text{for symmetric model} \\
                      J_{3} & \text{for generalized model} 
        \end{cases} 
\quad , \quad 
J_{zz} =  \begin{cases} J_{3} & \text{for symmetric model} \\
                       J_{3}/2 &  \text{for generalized model} 
        \end{cases}   \;  .  
\end{equation} 
\end{widetext}
When $J_{2}$ is sufficiently larger than $J_{3}$,
crystallization of upward spins (or, hardcore bosons) occurs 
and a `solid' phase $\ldots 333333 \ldots$ or $\ldots 666666 \ldots$
is realized.  Note that both states spontaneously break link-parity 
(translational symmetry is not broken).  
While this crystallization is obvious for the symmetric model,  
the situation is slightly subtle for the generalized model since 
the coupling between tetrahedra is $XY$-like.   
Fortunately, even in this case we can show that either of the two 
is selected by a weak $J_{3}$-coupling.   

In order to excite this `crystallized' ground state {\em magnetically}, 
a finite amount of energy ($\propto J_{2}+J_{3}$) has to be spent and
this leads to the type-I plateau at $m^{z}/m_{\text{sat}}=1/2$.
In particular, it is clear from the absence of
any kind of kinetic terms that the magnetization process for 
the symmetric model is step-like \cite{Honecker-M-T-00}  
under the conditions assumed here ($J_1 \gg J_2, J_3$). 

As is obvious from Fig.~\ref{fig:single-states}, for $J_{1}\approx 
J_{2}$ two different magnetic particles (`3' or `6' and `9') 
degenerate and competition between two different plateau phases 
(type-I and II) occurs.  
To investigate it, we can use a similar method to that used in 
section I for $J_2 \approx J_1$ and $J_{3}
\ll J_1, J_2$.  
Let us fix the magnetization $m^{z}/m_{\text{sat}}=1/2$.  
Then the half magnetized state is dominated by the three
types of tetrahedra triplets `3', `6' and `9' 
and we can construct an effective $S=1$ Hamiltonian using these
nearly degenerate states. We identify the tetrahedron states 
$\frac{1}{\sqrt{2}}(|\text{`3'}\rangle \pm |\text{`6'}\rangle )$ and
$|\text{`9'}\rangle$ with $S^{z}=\pm 1$ and $S^{z}=0$, respectively,
and obtain, in lowest non-trivial order, the following effective
Hamiltonian: 
\begin{align}
{\cal H}_{\text{eff-2}} &= -\frac{J_{3}}{4} \sum_{j} 
\left[ (\widetilde{S}_{j}^{x})^{2}-
(\widetilde{S}_{j}^{y})^{2}\right]
\left[ (\widetilde{S}_{j+1}^{x})^{2}
-(\widetilde{S}_{j+1}^{y})^{2}\right]  \notag \\
& +(J_{2}-J_{1}) \sum_{j} (\widetilde{S}_{j}^{z})^{2} \; . 
\label{eqn:eff_Ham_mag}
\end{align} 
It is clear that the above model has two phases: 
when $J_{3}$ is much larger than $J_{2}-J_{1}(>0)$, the ground state
is given by a product $\otimes_{j} \frac{1}{\sqrt{2}}
(|\text{`3'}\rangle_{j} \pm |\text{`6'}\rangle_{j} )$ while
the trivial product $\otimes_{j}|\text{`9'}\rangle_{j}$ becomes
the ground state when $(\widetilde{S}^{z})^{2}$ is dominant.

If we notice that the above model is in fact classical,
it is not difficult to know what kind of transition occurs
between two phases.
Although the Hamiltonian (\ref{eqn:eff_Ham_mag}) looks 
complicated, the fact that $  [ (\widetilde{S}^{x}_{j})^{2} 
- (\widetilde{S}^{y}_{j})^{2},
(\widetilde{S}_{j}^{z})^{2}]=0$ enables us to 
rewrite it in terms of classical spins
$\sigma_{j}$ which take three values $-1$, 0, and 1.
That is, if we identify $\frac{1}{\sqrt{2}}(|1\rangle \pm
|-1\rangle )$ and $|0\rangle$ with $\sigma = \pm 1$ and
$\sigma =0$ respectively and perform the replacement 
\begin{equation}
\sigma_{j} =  (\widetilde{S}^{x}_{j})^{2} 
- (\widetilde{S}^{y}_{j})^{2} \quad , \quad 
\sigma_{j}^{2} = (\widetilde{S}_{j}^{z})^{2} \;  ,
\end{equation}
the Hamiltonian (\ref{eqn:eff_Ham_mag}) reduces to
\begin{equation}
{\cal H}^{\prime}_{\text{eff-2}} = -\frac{J_{3}}{4}\sum_{j}
\sigma_{j}\sigma_{j+1} + (J_{2}-J_{1}) \sum_{j}\sigma^{2}_{j} \;  .
\label{eqn:Blume-Capel}
\end{equation}
This is nothing but the Blume-Capel model whose phase diagram is 
well known \cite{Cardy-96}; 
for sufficiently low temperatures (zero, in our case), 
there are a ferromagnetic phase with all $\sigma$ taking 1 or $-1$ and
the so-called vacancy phase where $\sigma=0$ at all sites and 
these two phases are separated by a first-order transition.       
In the language of the original quantum spins, the ferromagnetic 
state corresponds to $\otimes_{j} |\text{`3'}\rangle_{j}$ or 
$\otimes_{j} |\text{`6'}\rangle_{j}$ and 
the vacancy state to $\otimes_{j}|\text{`9'}\rangle_{j}$.   
Therefore, the aforementioned two plateau phases correspond to 
the ferromagnetic- and the vacancy phases, respectively.  

The above argument is based on the lowest-order effective Hamiltonian.
If we proceed to the next order, a new interaction appears:
\begin{equation}
\sum_{j} \sigma_{j}\sigma_{j+1}(\sigma_{j}-\sigma_{j+1}) \; .
\end{equation}
Fortunately, this type of interactions does not affect the two phases
mentioned above and we may expect that our result will remain 
qualitatively correct in higher orders of perturbation theory.  

\begin{figure}[b]
\begin{center}
\epsfig{file=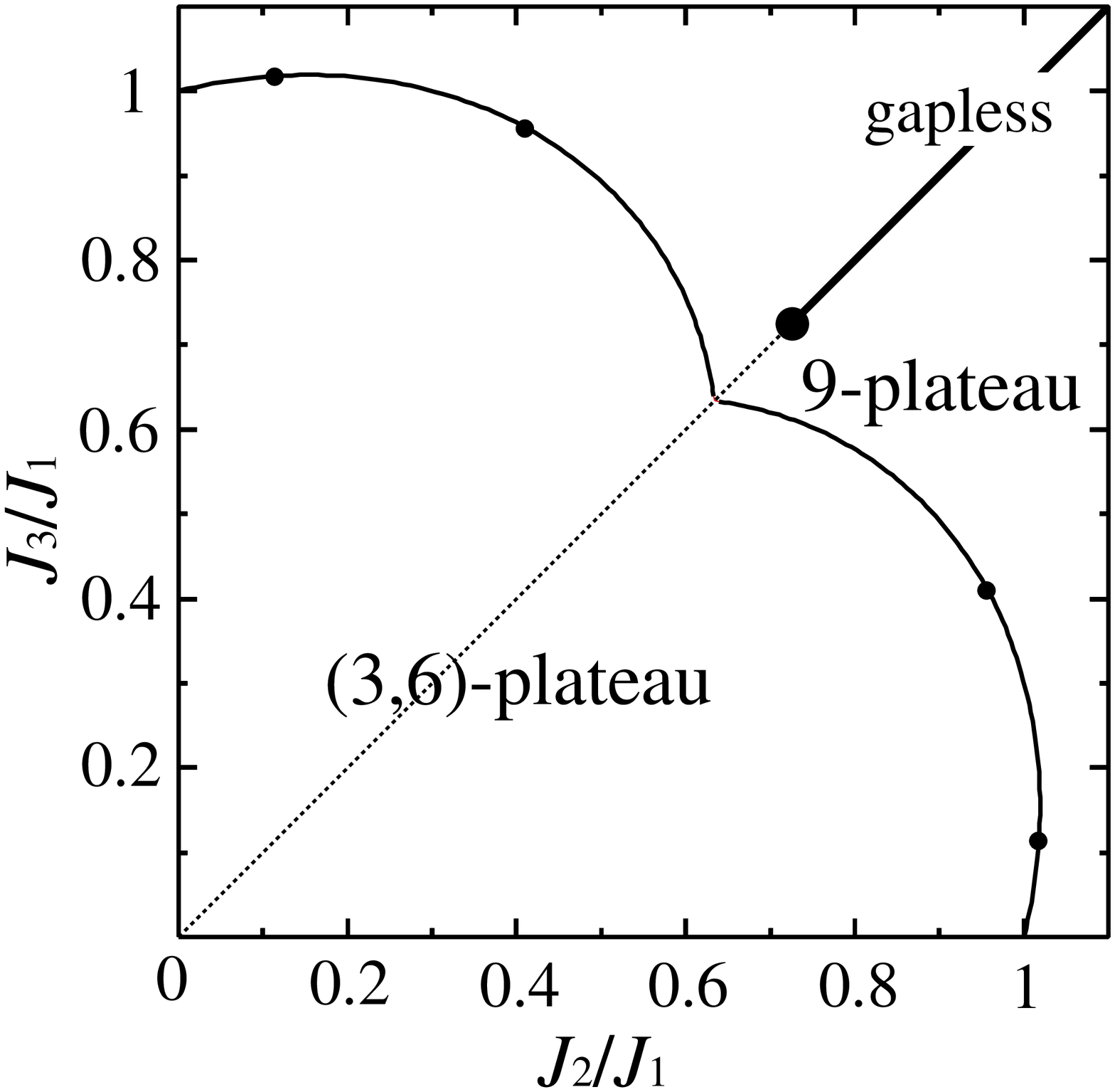,height=70mm}
\end{center}
\caption{Phase diagram for $m^{z}/m_{\text{sat}}=1/2$ obtained by 
low-order perturbation. Inside the boundary, (3,6)-plateau occurs.  
Transition points determined by a similar method to that used in 
section I are plotted by small dots (numerical data were taken 
from Ref.\onlinecite{Tonegawa-N-K-96}).  
A large dot on the symmetric line, which separates (3,6)-plateau- and 
non-plateau phases (shown by a thick line), was taken from 
Ref.\onlinecite{Honecker-M-T-00}.   
From the results for the bond-alternating $S=1$ chain 
\cite{Tonegawa-N-K-96,Totsuka-97}, 
we believe that the non-plateau phase is realized {\em only} on 
the symmetric line. 
Poor convergence around the symmetry axis $J_{2}=J_{3}$ may be 
attributed to the proximity to the criticality.}
\label{fig:phase-diag2}
\end{figure}

We determine the phase boundary by perturbation expansion. By
comparing the energies of both states obtained by perturbation expansion 
(up to second order), we obtain the following equation for the phase boundary
\begin{equation}
J_{1} - J_{2}+\frac{1}{4}J_{3} -\frac{53J_{3}^{2}}{64J_{2}} =0 \;  .
\end{equation}
The leading term $J_{1}-J_{2}+J_{3}/4$ coincides with that given by 
eq.(\ref{eqn:Blume-Capel}).    
In Fig.\ref{fig:phase-diag2} we show the phase diagram 
for finite magnetization 
$m^{z}/m_{\text{sat}}=1/2$.   
Inside the boundary, we have parity breaking states 
($\ldots 33333 \ldots$ or $\ldots 66666 \ldots$) similar to 
that found in Ref.\onlinecite{Totsuka-98} and 
a unique parity-symmetric 
state ($\ldots 99999 \ldots$) outside.    
\section{Summary and Discussion}
\label{section:summary}
Motivated by recent discovery of tellurates 
$\text{Cu}_{2}\text{Te}_{2}\text{O}_{5}\text{X}_{5}$ (X=Cl or Br), 
we considered low-lying excitations and the magnetization 
process of a system of coupled tetrahedra built of four spins 
$S = 1/2$.   

To investigate possible low-lying singlets and transitions 
between several ground states, we derived a classical Ising 
chain in a magnetic field as an effective Hamiltonian.   
Using it, we showed the growth of binding energy between 
singlet particles triggers a first-order transition between 
the simple rung-dimer (RD) phase and the spin-1 dimer (S1D) 
phase.   

Most of excitations in the RD phase are immobile 
or strongly localized due to the geometry (the so-called 
`orthogonal dimer' structure\cite{Miyahara-U-99}).   
On the other hand, in addition to immobile singlets, excitations with 
much larger mobility (with dispersion $\sim J_{3}$) 
are also allowed in the S1D phase.   
In both phases, there exists a `window' in the vicinity of 
the point $J_{1}=J_{2}$, where the lowest excitation is not a triplet,
but many singlet excitations populate the gap to the lowest triplet.   
A similar situation is known to occur  
for a frustrated ladder ($J_{2}=J_{3}=J_{4}$) \cite{Kotov-S-E-99}.    
In certain limits, these singlets can be viewed as bound states  
of dimer excitations.   
It would be worth mentioning here that these low-lying singlets 
in the S1D phase are {\em not} included in the Hilbert space of 
the effective spin-1 chain.   

We also determined the phase boundary by two different 
methods (effective Hamiltonian and a numerical method) 
and compared the results to obtain satisfactory agreement.   
In particular, for a generalized coupling $J_{3} \gg J_{4}$ 
we concluded that a weakly-coupled perfect 
($J_{1}=J_{2}$) tetrahedra is in the RD phase.    
As is suggested by this, the effect of couplings between 
tetrahedra (in particular, the resulting ground states) is 
sensitive to the detail of the couplings ({\em symmetric} 
or {\em generalized}, $J_{1}$ and $J_{2}$) and 
detailed information on the couplings 
is crucial in comparison with experiments.     

One way to distinguish between two phases (RD and S1D) experimentally 
would be to use optical probes.   
For example, the Raman operator which creates singlet excitations 
belonging to A-representation of $\text{S}_{4}$ is given 
essentially by ${\cal H}_{1}$ with $J_{3}=J_{4}$.     
Since it annihilates the RD ground state, the scattering intensity 
should be very weak for (cc)-polarization (c-axis is parallel to 
the chain axis), while it creates 
singlet combinations of two `$\gamma$' tetrahedra in the 
S1D phase.   
(The Raman operator corresponding to B-representation takes 
the form $\sum_{i}K (\bolS_{1,i}-\bolS_{3,i}){\cdot}
(\bolS_{2,i+1}-\bolS_{4,i+1})$, which excites elementary singlets 
and two-triplet bound states both in the RD- and the S1D phase.)     

In section III, we investigated the magnetization process.  
There appears a 1/2-plateau in the magnetization curve.  
In the dimer region $J_{1} \gg J_{2}, J_{3}$, 
the plateau is accompanied by discrete symmetry (parity) 
breaking and is attributed to the 
ordered state: $\ldots 3333333 \ldots$ or 
$\ldots 66666666 \ldots$, whereas 
in the `$S=1$' region a parity-symmetric 
state $\ldots 9999999 \ldots$ 
realizes.  Note that the translational symmetry is not 
broken at all.  
The transition between these two types of plateaus is described by 
a simple pseudo-spin ($S=1$) Hamiltonian equivalent to 
the classical Blume-Capel model.   

\begin{figure}[H]
\begin{center}
\epsfig{file=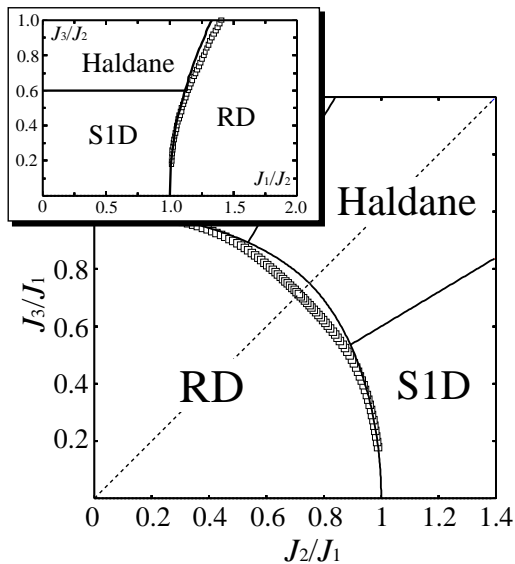,height=75mm}
\end{center}
\caption{Ground-state phase diagram for the symmetric model 
($J_{3}=J_{4}$).  
Note that the phase diagram is symmetric under $J_{2} \leftrightarrow 
J_{3}$.   
The phase boundary between RD- and Haldane phases as obtained from 
the effective Hamiltonian ${\cal H}_{\text{Ising}}$ 
(eq.\ref{eqn:effective1}) is shown by a full line and that from 
numerical data (for 16 sites, or $N=8$; Ref.~\onlinecite{Totsuka-N-H-S-95}) 
by open squares.  The inset shows the same data using the parametrization 
adopted in Ref.~\onlinecite{Brenig-B-01}.}
\label{fig:phase-diag1}
\end{figure}

Finally, we give some brief comments on the effects of possible 
three-dimensional couplings which recent experiments\cite{Lemmens-01} 
suggest are relatively large.   
We carried out preliminary calculation assuming the simplest 
interchain coupling ($J_{\perp}$) compatible with the crystal 
structure and found the following: 
(i) the ground state problem is again described by the classical 
(3D) Ising model at least for small couplings and 
(ii) the phase boundary between RD- and S1D phase is relatively 
insensitive to the interchain coupling $J_{\perp}$.  
Therefore we may expect that if we are in the S1D phase when
$J_{\perp}=0$ then so are we even for small but finite 
$J_{\perp} (\neq 0)$.  
However, new phenomena show up when the dynamics in the triplet 
sector is considered.  
An analyses analogous to the one presented in section II B 
suggests that the triplet (`9') gap gets reduced substantially 
while the singlet-singlet 
gap slightly increases as we increase $J_{3}$ and $J_{\perp}$; 
spin gaps in the {\em effective} spin-1 sector finally close and 
three-dimensional antiferromagnetic ordering may take place in the 
effective spin-1 system (not in the original spin-1/2 system).  
A similar phenomenon has been used as a theoretical trick in the context of 
the so-called composite-spin models \cite{Timonen-L-85,Koga-K-00}.    
The region of the above ordered phase blows up from the transition line 
between S1D and Haldane (see Fig.\ref{fig:phase-diag1}) and grows 
as $J_{\perp}$ is increased.   
\begin{acknowledgments}
This work was supported by the German Federal Ministry for 
Education and Research (BMBF) under the contract 03MI5HAN5.  
The hospitality and support of Hahn-Meitner-Institut Berlin, 
where this work was completed, is gratefully acknowledged.  
The authors thank W.~Brenig, A.~Harrison, F.~Mila, Y.~Nishiyama, 
and T.~Vecua for discussions.  
They are also grateful to P.~Lemmens and P.~Millet 
for discussions and helpful correspondences.   
\end{acknowledgments}

\begin{table}[h]
\begin{center}
\begin{tabular}{l c c c c} \hline \hline
\makebox[15mm]{phase} & \makebox[50mm]{energy} & 
\makebox[20mm]{spin} & 
   \makebox[25mm]{degeneracy} & 
\makebox[20mm]{symmetry at $\Gamma$}  \\ \hline 
RD & 
$\Delta_{\text{sing}}^{\text{RD}}$ (see eq.(\ref{eqn:RD-sing})) 
& singlet & $N$ & B \\
 & $\Delta_{\text{sing}}^{\text{RD}} +2(n-1)|h_{\text{IM}}|$ & 
singlet & $N$ & A ($n$:even) B ($n$:odd) \\  
 & $2N |h_{\text{IM}}|$  & singlet & 1 \\
 & $J_{1}$ & triplet & $N$ & E \\
 & $2J_{1}-J_{2,3}$ & triplet & $N$ & A \\ 
\hline 
S1D & $\Delta^{\text{sing}}_{\text{S1D}}$ & singlet & $N$ & B \\
 &  $\Delta^{\text{sing}}_{\text{S1D}}+2(n-1)h_{\text{IM}}$ 
& singlet & $N$ & A ($n$:even) B ($n$:odd) \\
 &  $\Delta^{\text{triplet}}_{\text{type-1}}$ & 
triplet & $N$ & E \\
 &  $\Delta^{\text{triplet}}_{\text{type-2}}$ & 
triplet & non-degen. band & B 
\\ \hline\hline
\end{tabular}
\end{center}
\caption{Energy, spin, degeneracy, and symmetry classification 
(in terms of $\text{S}_{4}$) of typical excitations.  
$N$ denotes the number of tetrahedra.}
\end{table}

\begin{thebibliography}{10}
\bibitem{Lhuillier-S-F-00}
C.~Lhuillier, P.~Sindzingre, and J-B.~Fouet. 
\newblock {\tt cond-mat/0009336}; \\
R.~Moessner. {\tt cond-mat/0010301}. 

\bibitem{Gelfand-91}
M.P.~Gelfand. 
\newblock Phys.Rev.B {\bf 43}, 8644 (1991);\\
B.~Sutherland. 
\newblock Phys.Rev.B {\bf 62}, 11499 (2001)

\bibitem{Bose-G-93}
I.~Bose and S.~Gayen,  
\newblock Phys.Rev.B {\bf 48}, 10653 (1993);\\
A.K.~Kolezhuk and H-J.~Mikeska. 
\newblock Int.J.Mod.Phys.B, {\bf 12}, 2325 (1998).

\bibitem{Johnsson-T-M-M-00}
M.~Johnsson, K.W.~T\"{o}rnroos, F.~Mila, and P.~Millet. 
\newblock Chem.Mater., {\bf 12}, 2853 (2000). 

\bibitem{Brenig-B-01}
W.~Brenig and K.W.~Becker. 
\newblock Phys.Rev.B, {\bf 64}, 214413 (2001).  

\bibitem{Lemmens-01}
P.~Lemmens et al.  
\newblock Phys.Rev.Lett., {\bf 87}, 227201 (2001).  

\bibitem{Xian-95} 
Y.~Xian. 
\newblock Phys.Rev.B, {\bf 52}, 12485 (1995); \\
X.~Wang. 
\newblock cond-mat/9803290. 

\bibitem{Haldane-83}
F.D.M.~Haldane.
\newblock Phys.Rev.Lett., {\bf 50}, 1153 (1983).

\bibitem{Yamamoto-95}
Y.~Kato and Y.~Tanaka.
\newblock J.Phys.Soc.Jpn., {\bf 63}, 1277 (1994); \\
S.~Yamamoto. 
\newblock Phys.Rev.B, {\bf 52}, 10170 (1995).  

\bibitem{Totsuka-N-H-S-95}
K.~Totsuka, Y.~Nishiyama, N.~Hatano, and M.~Suzuki.
\newblock J.Phys.condensed matter, {\bf 7}, 4895 (1995). 

\bibitem{denNijs-R-89}
M.~den Nijs and K.~Rommelse.  
\newblock Phys.Rev.B {\bf 40}, 4709 (1989); \\
S.M.~Girvin and D.~Arovas.  
\newblock Phys.Scr.T, {\bf 27}, 156 (1990).  

\bibitem{Note-PD}
Actually, we carried out cluster expansion up to $(J_{3}/J_{2})^{5}$ 
to find that a tiny descrepancy between analytic- and numerical 
results was removed by including the third-order correction.    

\bibitem{Mambrini-T-M-99}
M.~Mambrini, J.~Tr\'{e}bosc, and F.~Mila. 
\newblock Phys.Rev.B, {\bf 59}, 13806 (1999). 

\bibitem{Waldtmann-98}
P.~Lecheminant et al. 
\newblock Phys.Rev.B, {\bf 56}, 2521 (1997); \\
C.~Waldtmann et al. 
\newblock Euro.Phys.J.B, {\bf 2}, 501 (1998).  

\bibitem{Ramirez-H-W-00}
A.P.~Ramirez, B.~Hessen, and M.~Winklemann. 
\newblock Phys.Rev.Lett., {\bf 84}, 2957 (2000); \\
P.~Sindzingre et al. 
\newblock {\em ibid.} {\bf 84}, 2953 (2000). 

\bibitem{Kotov-S-00}
V.N.~Kotov and O.P.~Sushkov. 
\newblock Phys.Rev.B, {\bf 61}, 11820 (2000).  

\bibitem{Kotov-O-S-W-00}
V.N.~Kotov, J.~Oitmaa, O.~Sushkov, and Z.~Weihong. 
\newblock Phil.Mag.B, {\bf 80}, 1483 (2000).  

\bibitem{Kotov-Z-S-01}
V.N.~Kotov, M.E.~Zhitomirsky, and O.P.~Sushkov. 
\newblock Phys.Rev.B, {\bf 63}, 064412 (2001). 

\bibitem{note-1}
This localization must be closely related to the non-existence 
of the so-called $S=1/2$ edge states in the $S=1$ dimer 
phase \cite{Yamamoto-95}.   
If we weaken dimerization to enter the Haldane phase, 
the edge states show up and this type 
of excitations will dissociate into two $S=1/2$ objects.   

\bibitem{Mattis-book}
D.C.~Mattis. 
\newblock {\em The theory of magnetism}, Springer-Verlag, 1981. 

\bibitem{note-2}
For example, in the decoupling limit $J_{3}=0$ we can trace a level 
corresponding to a `$\alpha$' tetrahedron  
as we cross $J_{2}=J_{\text{2,c}}
=1$.   When $J_{3}>0$, however, an excited state corresponding to 
one `$\alpha$' tetrahedron in the RD background and that  
in the sea of `2' tetrahedra should be considered quite different 
from each other.  

\bibitem{Honecker-M-T-00}
A.~Honecker, F.~Mila, and M.~Troyer. 
\newblock Euro.Phys.J. B {\bf 15}, 227 (2000). 

\bibitem{Cardy-96}
see for example, J.L.~Cardy.
\newblock {\em Scaling and renormalization in statistical 
physics}, Cambridge University Press, 1996.  

\bibitem{Okazaki-M-S-00}
N.~Okazaki, J.~Miyoshi, and T.~Sakai. 
\newblock J.Phys.Soc.Jpn. {\bf 69}, 3 (2000). 

\bibitem{Totsuka-98}
K.~Totsuka. 
\newblock Phys.Rev.B {\bf 57}, 3454 (1998).     

\bibitem{Tonegawa-N-K-96}
T.~Tonegawa, M.~Nakao, and M.~Kaburagi. 
\newblock J.Phys.Soc.Jpn. {\bf 65}, 3317 (1996).

\bibitem{Totsuka-97}
K.~Totsuka.
\newblock Phys.Lett.A {\bf 228}, 103 (1997).

\bibitem{Miyahara-U-99}
S.~Miyahara and K.~Ueda. 
\newblock Phys.Rev.Lett. {\bf 82}, 3701 (1999); 
K.~Totsuka, S.~Miyahara, and K.~Ueda. 
\newblock {\em ibid} {\bf 86}, 520 (2001).  

\bibitem{Kotov-S-E-99}
V.N.~Kotov, O.P.~Sushkov, and R.~Eder.  
\newblock Phys.Rev.B {\bf 59}, 6266 (1999).   

\bibitem{Timonen-L-85}
J.~Timonen and A.~Luther. 
\newblock J.Phys.C {\bf 18}, 1439 (1985); \\
J.~Solyom and J.~Timonen. 
\newblock Phys.Rev.B {\bf 40}, 7150 (1089).

\bibitem{Koga-K-00}
A.~Koga and N.~Kawakami. 
\newblock Phys.Rev.B {\bf 61}, 6133 (2000).
\end{thebibliography}
\end{document}